# Multifunctional *in vivo* vascular imaging using near-infrared II fluorescence

Guosong Hong[1,3], Jerry C. Lee[2,3], Joshua T. Robinson[1], Uwe Raaz[2], Liming Xie[1], Ngan F. Huang[2], John P. Cooke[2] & Hongjie Dai[1]

[1] Department of Chemistry, Stanford University, Stanford, California 94305, USA

[2] Division of Cardiovascular Medicine, School of Medicine, Stanford University, Stanford, California 94305, USA

[3] These authors contribute equally to this work.

Correspondence should be addressed to: hdai@stanford.edu, jcooke@stanford.edu and ngantina@stanford.edu

## Abstract

*In vivo* real-time epifluorescence imaging of mouse hindlimb vasculatures in the second near-infrared region (NIR-II, 1.1–1.4 μm) is performed using single-walled carbon nanotubes (SWNTs) as fluorophores. Both high spatial resolution (~30 μm) and temporal resolution (<200 ms/frame) for small vessel imaging are achieved 1~3 mm deep in the tissue owing to the beneficial NIR-II optical window that affords deep anatomical penetration and low scattering. This spatial resolution is unattainable by traditional NIR imaging (NIR-I, 0.75–0.9 μm) or microscopic computed tomography (micro-CT), while the temporal resolution far exceeds scanning microscopic imaging techniques. Arterial and venous vessels are unambiguously differentiated using a dynamic contrast-enhanced NIR-II imaging technique based on their distinct hemodynamics. Further, the deep tissue penetration, high spatial and temporal resolution of NIR-II imaging allow for precise quantifications of blood velocity in both normal and ischemic femoral arteries, which are beyond the capability of ultrasonography at lower blood velocity.

## Introduction

Development of new therapies for peripheral arterial diseases (PADs) may be facilitated by imaging that provides anatomic and hemodynamic information with high spatial and temporal resolution. However, current methods for assessing vasculature and hemodynamics in small vessels *in vivo* are suboptimal.[1] For imaging vascular structures, microscopic computed tomography (micro-CT) and magnetic resonance imaging (MRI) can resolve features down to ~100 μm with deep penetration, but are limited by long scanning and post-processing time and difficulties in assessing vascular hemodynamics.[2,3] On the other hand, vascular hemodynamics are usually obtained by Doppler measurements of micro-ultrasonography with high temporal resolution of up to 1000 Hz, but spatial resolution attenuates with increased depth of penetration.[4]

*In vivo* fluorescence-based optical imaging has inherent advantages over tomographic imaging owing to high temporal and spatial resolution.[5,6] Single-walled carbon nanotubes (SWNTs), nanoscale cylinders of rolled-up graphite sheets comprised of carbon, are an emerging nanomaterial with unique optical properties for *in vivo* anatomical imaging,[7,8] tumor detection[9,10] and photothermal treatment.[10-12] One unique feature of SWNTs is their intrinsic fluorescence in the second near-infrared (NIR-II, 1.1–1.4 μm) window upon excitation in the traditional near-infrared region (NIR-I, 0.75–0.9 μm) with large Stokes shift up to ~400 nm. Compared to the NIR-I window that has been extensively explored for *in vitro* and *in vivo* imaging,[13-18] the longer wavelength emission in NIR-II makes SWNTs advantageous for imaging owing to reduced photon absorption[19,20] and scattering by tissues,[8,21] negligible tissue autofluorescence and thus deeper tissue penetration,[7,8,22] allowing for unprecedented fluorescence-based imaging resolution of anatomical features deep to the skin.

Here we report the use of biocompatible, brightly fluorescent SWNTs as NIR-II imaging agents for imaging vascular structures down to ~30 μm in mouse hindlimb using an epifluorescence imaging method with an indium-gallium-arsenide (InGaAs) imaging system. Compared to micro-CT, NIR-II fluorescence imaging attains a ~3-fold improvement in spatial resolution. Further, NIR-II imaging allows for differentiation of arteries from veins through principal component analysis (PCA)[8,23] to obtain dynamic contrast-enhanced imaging. We can also quantify femoral artery blood flows in both normal and ischemic hindlimbs, and reveal the degree of occlusion due to ischemia. Thus, a single NIR-II imaging modality enables multi-functional imaging capable of accomplishing what is typically done by several traditional techniques including micro-CT, ultrasonography and MRI, making it a unique method that incorporates many desirable features such as high spatial resolution (~30 μm), fast acquisition (<200 ms), good tissue penetration (1~3 mm), vessel specification and blood flow quantification.

## Results

**NIR-I and NIR-II fluorescence imaging of vasculatures.** First, to glean the differences of *in vivo* NIR-I and NIR-II fluorescence imaging, we made biocompatible SWNT-IRDye-800 conjugates as dual-color imaging agents, where IRDye-800 (IRDye 800CW, LI-COR) was a commercial NIR-I fluorophore. High-pressure carbon monoxide conversion (HiPCO) SWNTs were stably suspended by biocompatible surfactants of 75% DSPE-mPEG (5kDa) and 25% DSPE-PEG(5kDa)-$NH_2$ with amine groups covalently functionalized with IRDye-800 (**Figs. 1a** and **S1a-c**). Both the SWNT and IRDye-800 label could be excited by a 785 nm laser, but exhibited different emissions. The IRDye-800 dye emitted at ~800 nm in the NIR-I, while the SWNT emitted in the 1.1–1.4 μm NIR-II region (**Fig. 1b**). The dual-color emission of SWNT-IRDye-800 conjugate ensured co-localization of SWNTs and IRDye-800 dyes, and enabled us to image the same tissue in two distinct spectral windows so as to evaluate the performance of photons in different wavelengths for live animal imaging.

For parenteral administration and live animal imaging, we prepared and injected a solution (200 μL) of 0.10 mg·$mL^{-1}$ (1.0 mg·$kg^{-1}$ body weight) SWNT-IRDye-800 conjugates (**Fig. 1c**) into a nude mouse intravenously. We estimated the maximum SWNT concentration in the blood would be 17× lower than the half maximal inhibitory concentration (IC50) of vascular endothelial cells (**Fig. S1d**). The circulation half-time of DSPE-mPEG functionalized SWNTs was ~5 h,[8] and our previous studies had shown the lack of acute or long-term toxicity of such PEGylated SWNTs *in vivo*.[10,24,25] The mouse was illuminated using a 785 nm laser at 8 mW·$cm^{-2}$ (**Fig. 1d**) and imaged in the NIR-I (**Fig. 1e–g**) using a Si camera and in the NIR-II (**Fig. 1h–j**) using an InGaAs camera equipped with different emission filters. An adjustable lens set was used to obtain three different magnifications: 1× (whole body), 2.5× (entire hindlimb) and 7× (partial hindlimb).

We found that all images in the NIR-I region employing IRDye-800 fluorescence manifested indistinct vascular anatomy. The corresponding cross-sectional intensity profiles were characterized by broad peaks (**Fig. 1e–g**, bottom). Presumably, these characteristics were related to significant scattering and absorbance of NIR-I photons, which limited the depth and resolution of traditional NIR imaging *in vivo*.[8] In contrast, when the same mouse was imaged in the NIR-II region by detecting SWNT fluorescence, there was a substantially improved spatial resolution of vessels at all magnifications. Moreover, the NIR-II window clearly visualized smaller, higher-order branches of blood vessels at higher magnifications. Cross-sectional intensity profiles all exhibited sharp peaks, with the calculated vessel diameter values consistent with expected values (**Fig. 1h–j**, bottom). In contrast, it was impossible to calculate vessel diameters based on the NIR-I images, in which we observed a 2~3-fold broadening of the cross-sectional profiles on average.

**NIR-II and micro-CT for vessel imaging.** Micro-CT is a commonly-used three-dimensional (3D) X-ray imaging technique based on tomographic reconstruction with spatial resolution down to ~100 μm and excellent penetration depth.[26,27] Accordingly, we compared the spatial resolution of the proximal femoral artery and vein in the same mouse, achieved by NIR-II and micro-CT methods (**Fig. 2a,b**). A cross-sectional analysis was performed across the blood vessels in each image, and the corresponding intensity profiles are shown for NIR-II image (**Fig. 2c**) and micro-CT image (**Fig. 2d**). Two peaks were identified in each plot, corresponding to femoral artery and vein, and were fitted into two Gaussian functions to extract the widths. The vessel widths extracted from NIR-II image (0.284 mm and 0.255 mm) agreed with those from micro-CT image (0.292 mm and 0.247 mm). The analysis of micro-CT images validated the vessel diameters measured by NIR-II imaging, suggesting the two techniques are comparable in imaging vascular structures ~100s μm in diameter.

To compare the resolution limits between NIR-II and micro-CT, we determined the smallest vessels these two techniques were able to discern. In the distal hindlimb of the mouse, the NIR-II image showed greater numbers of small vessels compared to the micro-CT image at the same location (**Fig. 2e,f**). The smallest measurable vessel by NIR-II had a Gaussian-fit diameter of only 35.4 μm (**Fig. 2g**), while micro-CT could not discern any vessel smaller than ~100 μm in diameter (**Fig. 2h**). Furthermore, NIR-II method generated images much faster than micro-CT (300 ms for NIR-II and 2 h for micro-CT).

**Differentiation of arterial and venous vessels**. To determine whether NIR-II fluorescence imaging could distinguish arterial from venous circulation, we monitored the blood flow inside the vessels by recording videos in the NIR-II window immediately upon tail-vein injection of SWNT fluorophores. To enable greater temporal resolution (i.e., shorter exposure time) for dynamic recordings, we used an 808 nm laser at 140 mW·cm$^{-2}$ for more efficient excitation of SWNT fluorophores.

When a 200 μL solution containing 0.10 mg·ml$^{-1}$ biocompatible SWNTs was injected into the tail vein of a nude mouse (Mouse C1), we observed an NIR-II signal from the SWNTs in the proximal femoral artery within 5 s, and the entire femoral artery and some of the proximal musculature after ~8 s (**Fig. 3a,b**). Outflow of SWNTs into the femoral vein was reflected by the increased feature width (vascular bundle in the femoral sheath) at later time points but difficult to distinguish from the femoral artery by visual inspection due to their proximity (**Fig. 3c**). However, due to the time delay in the first appearance of the signal in the artery to its later appearance in the vein (Movie S1), it was possible to employ PCA to differentiate the two types of vessels.[8,23] Essentially, the PCA approach assigns image pixels to groups (components) based on their variance, i.e., pixels that vary similarly in time. We applied PCA to the video-rate images and clearly discriminated the arterial component from the venous component. By convention we color-coded the arterial PCA component in red and the venous component in blue (**Fig. 3d**). Similar results were obtained with an additional three mice (**Figs S2a-f**).

We repeated these studies in mice after surgically inducing unilateral hindlimb ischemia (Mouse I1–3). In this model, ligation and excision of the proximal superficial femoral artery and ligation of the deep femoral artery reduced limb perfusion by ~80% in the immediate postoperative period.[28] On the first postoperative day, we studied mice with SWNT-assisted NIR-II fluorescence imaging. After tail vein injection of SWNTs, we observed a marked delay in the appearance of fluorescent signals in ischemic limbs (**Fig. 3e-g**) compared to healthy limbs. As in normal animals, we performed PCA on the first 200 frames (~37.5 s) of video-rate recording (**Movie S2**) of the ischemic animal. However, due to the markedly reduced perfusion, venous return in the ischemic limb could not be observed at this time point (**Figs. 3h** and **S2g-j**). It was necessary to continue imaging for an extended period (>2 min) to visualize the femoral vein (**Fig. S3** and **Movie S3**). This delay in venous return was consistent with severe limb ischemia.

We also applied the dynamic contrast PCA approach to demonstrate that arteries and veins subserving a larger region of tissue could be distinguished. In these studies, we imaged a larger field of view including both the abdominal and femoral regions. Within 5 s of tail vein injection (**Movie S4**), signals were observed in the aorta as well as the iliac, femoral and epigastric arteries and their branches in the abdomen and pelvis (**Fig. 4b**). The NIR-II signal from these vessels peaked and then subsided by ~10 s post injection with increasing intensity in the tissue (**Fig. 4c**). Subsequently, signals were observed in veins draining these regions, including the previously identified femoral vein and hypogastric veins (**Fig. 4d**). As before, we used PCA to assign pixels to arterial or venous conduits based on their time variance (**Fig. 4e**), where arteries and veins were resolved. Notably, even the aorta can be seen in supine position of the mouse, indicating a penetration depth of >5 mm for NIR-II imaging. This result confirmed the capacity of NIR-II dynamic contrast-enhanced imaging to distinguish arteries from veins.

**Blood flow quantification in ischemic and control limbs.** NIR-II imaging revealed a dramatic delay in the appearance of fluorescence signal in ischemic hindlimbs versus control healthy limbs. We then assessed blood flow in the ischemic hindlimb of Mouse I1 quantitatively. From 12 s to 19.5 s post injection of SWNTs, propagation of the signal could be visualized from the proximal site of arterial occlusion to the distal femoral artery (**Fig. 5a**), presumably filling through collateral channels. The position of the signal front at each time point was then extracted from each frame and plotted against time, showing a linear relationship of flow distance vs. time with a blood velocity of 0.163 cm·s$^{-1}$ (**Fig. 5b**). To calculate blood flow, we measured the mean diameter of the femoral artery (174 μm). The measurement of blood flow velocity and arterial diameter permitted the calculation of femoral artery blood flow (2.33x10$^{-3}$ mL·min$^{-1}$).

We found the progression of NIR-II signal through the femoral artery in normal mice was so rapid that our current video imaging rate lacked sufficient temporal resolution. Alternatively, to assess blood velocity in these animals, we measured the increase in signal intensity in a region of interest (ROI) (a pre-defined segment of the femoral artery from PCA). We normalized the signal intensity to compensate for differences in actual injected dose or fluorescence quantum yield, and plotted the average ROI intensity vs. time (see Methods). In the animal with hindlimb ischemia, we observed an NIR-II intensity increase of 2.18%·s$^{-1}$ (**Fig. 5c**). Blood flow and NIR-II intensity were then plotted against each other (**Fig. 5d**) to yield a linear slope of ~0.0737 cm·%$^{-1}$ that correlated velocity with intensity change. We used this coefficient to convert the NIR-II intensity increase rate to blood velocity, in the normal animals where blood velocity exceeded the temporal resolution of our current imaging device. To further establish this translation coefficient, we reproduced the blood velocity quantification with two other ischemic mice, I2 and I3, and obtained an average value of 0.0747±0.0019 cm·%$^{-1}$ (**Fig. S4** and **Table 1**).

The intensity-to-velocity conversion coefficient was applied to measurements of femoral artery blood flow in a healthy, control mouse, where the blood velocity was so rapid that our current NIR-II imaging device was unable to track a discrete flow front over time (**Fig. 5e**). Alternatively we measured the NIR-II intensity change in an ROI of the femoral artery as a function of time (**Fig. 5f**), revealing a normalized NIR-II intensity increase of 68.7±5.2%·s$^{-1}$ before it peaked. By using the conversion coefficient of 0.0747 cm·%$^{-1}$, we were able to translate the intensity increase to a blood velocity of 5.13±0.39 cm·s$^{-1}$. To validate this measure of blood velocity, we used ultrasonography to obtain a Doppler-derived velocity in the same mouse at the same ROI of the femoral artery, by an operator blinded to the SWNT imaging values. In the same animal, the ultrasound measurement provided a Doppler-derived velocity of 4.97±0.17 cm·s$^{-1}$ (**Fig. 5g**), which was in good agreement with the NIR-II video-imaging result (a deviation of ~3%).

To further validate our intensity-to-velocity conversion coefficient and to understand the physics behind it, we used a simplified fluid dynamic system, in which SWNT solution was pumped into catheter tubing filled with purified water (**Fig. S5k**), and derived the coefficient based on NIR-II intensity increase (see Methods). By changing the experimental settings one at a time, we found the coefficient remained invariant (0.0764±0.0025 cm·%$^{-1}$) as we varied the injected SWNT concentration, SWNT fluorescence quantum yield, tubing diameter, ROI length and velocity. It did vary with the distance between the injection site and ROI (pre-ROI length) (**Fig. S5a-j** and **Table 2**). This finding was also confirmed by numerical simulations (**Fig. S5l-s** and **Table 2**) based on a linear flow model with axial mixing.[29] In the context of *in vivo* blood velocity quantification, since the pre-ROI length simply reflected the length of blood vessels in which the injected SWNTs had to travel before reaching the femoral artery, the coefficient should be invariant given the same type of animals with roughly same blood volume and therefore applicable from ischemic to control mice.

To demonstrate the reproducibility of the NIR-II fluorescence-based blood velocity quantification, we compared the Doppler-derived velocities to those of the NIR-II method in three additional control animals (**Fig. S6**). A summarized plot suggested excellent agreement of results between the two methods (**Fig. 5h**). Subsequently, we performed these studies in three ischemic mice. The marked reduction in blood velocity in ischemic limbs (**Fig. 5i**) measured by NIR-II method was consistent with previous reports.[28,30,31] It was noteworthy that blood velocity in the ischemic animals was low and beyond the detection capability of ultrasonography. Clearly, the NIR-II imaging technique proved advantageous over ultrasound by providing a broader dynamic range of blood velocity measurements *in vivo*.

## Discussion

Current methodologies for physiological imaging of PADs are suboptimal in that no single modality provides adequate spatial and temporal resolution to accurately assess all critical parameters, i.e. vascular structure, arterial inflow, venous outflow, and tissue perfusion. NIR-II imaging technique simultaneously provides anatomical and hemodynamic information and surpasses the need to use multiple imaging modalities to obtain equivalent data, owing to reduced tissue scattering and deeper anatomical penetration of NIR-II over shorter wavelengths. This is due to the inverse wavelength dependence ($\sim\lambda^{-1.1}$) of photon scattering as they travel through subcutaneous tissue and skin.[8]

NIR-II imaging compares favorably to micro-CT. Even with voxel dimensions of 40 μm, micro-CT was only capable of resolving vessels of ~100 μm, while NIR-II could distinguish vessels ~3× smaller, without obvious background signal from soft tissue. In terms of imaging depth, NIR-II can achieve a penetration depth of millimeters *in vivo* without losing fidelity of the vessel structures, whereas micro-CT is able to reconstruct the whole-body structure in 3D owing to the unlimited penetration of X-rays. The time required for micro-CT (hours) as opposed to NIR-II imaging (sub-second) also means longer anesthesia time and high radiation doses, carrying risks of nephrotoxicity, anaphylaxis, and tissue injury.[32,33]

NIR-II imaging has three salient advantages over ultrasonography. First, NIR-II is able to image smaller vessels. Even with a high frequency (40 MHz) transducer, the diameter of mouse vessels could not be accurately determined with ultrasound, due to poor spatial resolution and low contrast. Second, NIR-II imaging is able to resolve both arterial and venous vessels anatomically and hemodynamically using PCA. Third, NIR-II imaging can be used to acquire hemodynamic data in conditions of reduced flow (e.g., ischemic hindlimb), below the detection limit of ultrasonography.

Due to the many benefits of NIR-II fluorescence imaging over other pre-clinical imaging modalities, this dual-modality method may be useful in a variety of cardiovascular models.

Although our current work focused on assessing blood flow of small vessels, NIR-II imaging could be used to characterize *in vivo* the degree of stenosis or aneurysmal dilation of vessels. Because of its temporal resolution, it might also be useful for imaging dynamic changes in vessel diameter due to active vasodilation or vasoconstriction. It is conceivable that NIR-II fluorescence imaging with SWNTs and other novel NIR-II fluorophores including quantum dots[34,35] and synthetic organic dyes[36] could lead to many translational and clinical applications.

## Acknowledgements

This study was supported by grants from the National Cancer Institute of US National Institute of Health to H.D. (5R01CA135109-02), National Heart, Lung and Blood Institute of US National Institute of Health to J.P.C. (U01HL100397, RC2HL103400) and N.F.H. (K99HL098688), and a Stanford Graduate Fellowship to G. H.

## Author contributions

H.D., J.P.C., N.H., G.H. and J.C.L. conceived and designed the experiments. G.H., J.C.L., J.T.R., U.R., L.X. and N.H. performed the experiments. G.H., J.C.L., U.R., L.X., N.H., J.P.C. and H.D. analysed the data and wrote the manuscript. All authors discussed the results and commented on the manuscript.

## Additional information

This paper is now published online at *Nature Medicine*.

# Figures

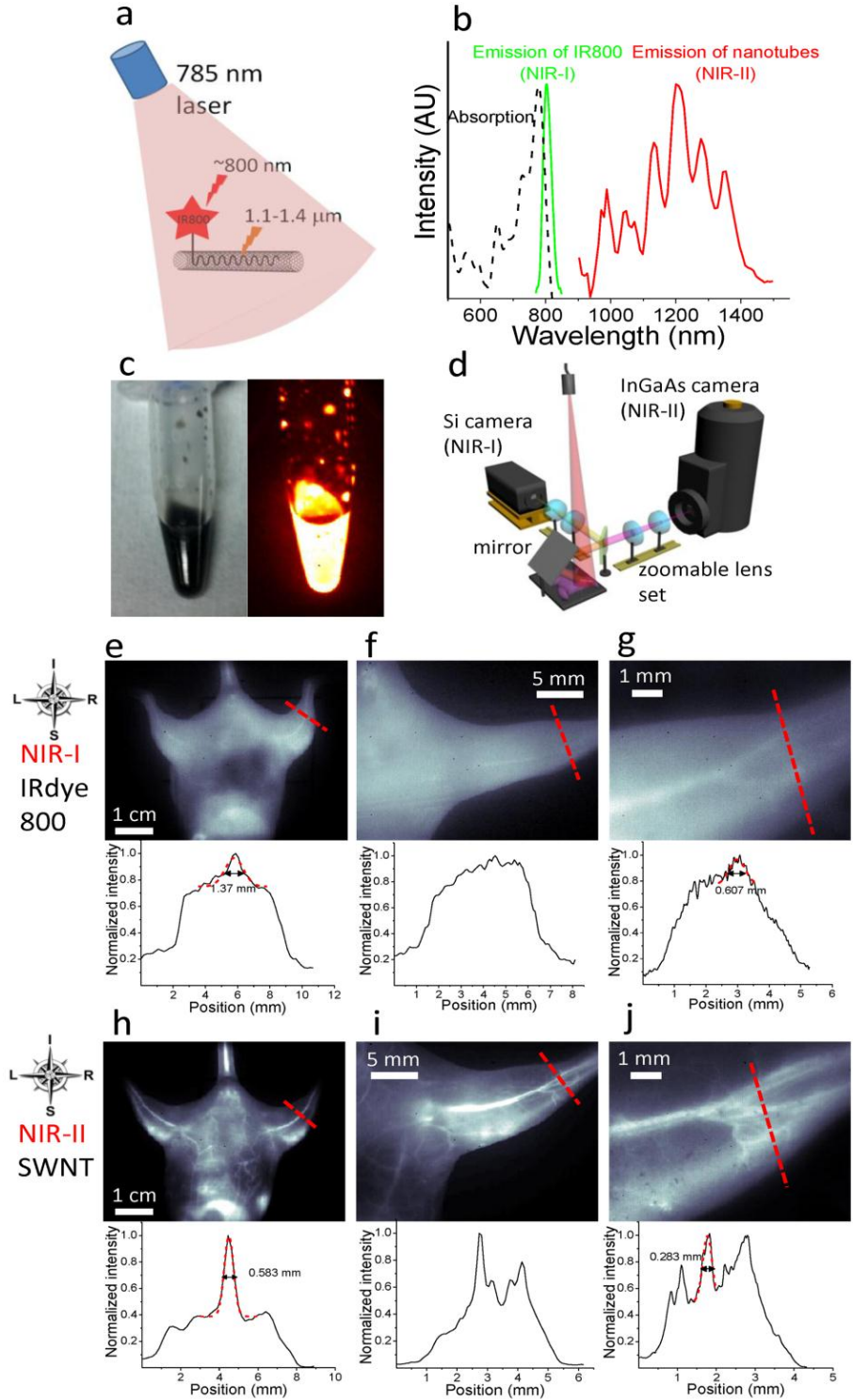

**Figure 1** NIR-I and NIR-II fluorescence imaging of blood vessels in the mouse. (**a**) A schematic showing that upon excitation of a 785 nm laser, the SWNT-IRDye-800 conjugate emits at the ~800 nm NIR-I

region from IRDye-800 dye and the 1.1−1.4 μm NIR-II region from the SWNT backbone. (**b**) Absorption spectrum of the SWNT-IRDye-800 conjugate (black dashed curve), emission spectrum of IRDye-800 dye (green solid curve) and SWNTs (red solid curve). (**c**) A digital camera photograph (left) and an NIR-II fluorescence image of injected solution containing 0.10 mg·ml$^{-1}$ SWNT-IRDye-800 conjugates. (**d**) Schematic of the imaging setup for simultaneous detection of both NIR-I and NIR-II photons using Si and InGaAs cameras. A zoomable lens set was used for adjustable magnifications. (**e–g**) NIR-I fluorescence images (top) and cross-sectional fluorescence intensity profiles (bottom) along red-dashed bars of a mouse injected with the SWNT-IRDye-800 conjugates. Gaussian fits to the profiles are shown in red dashed curves. (**h–j**) NIR-II fluorescence images (top) and cross-sectional fluorescence intensity profiles (bottom) along red-dashed bars of a mouse injected with the SWNT-IRDye-800 conjugates. Gaussian fits to the profiles are shown in red dashed curves.

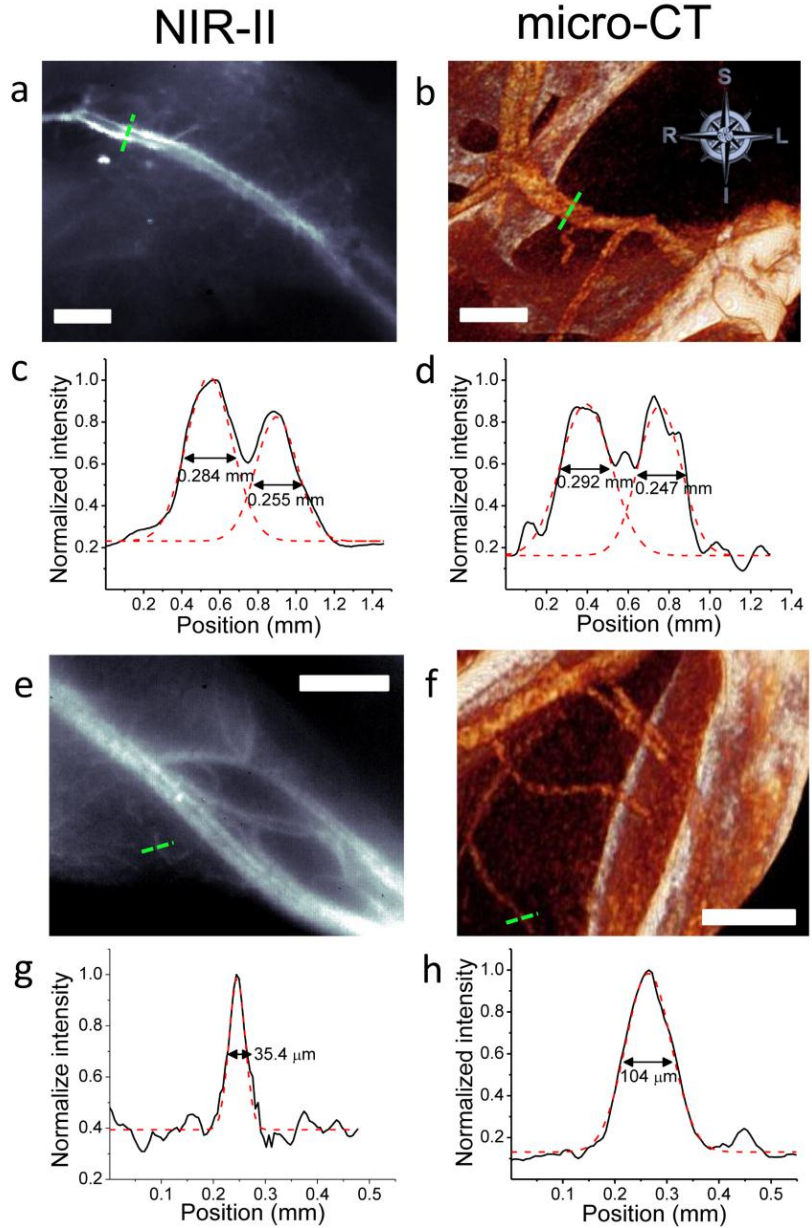

**Figure 2** NIR-II fluorescence and micro-CT imaging of hindlimb blood vessels. (**a**) A NIR-II SWNT fluorescence image of a mouse thigh. (**b**) A micro-CT image showing the same area of the thigh as in **a**. (**c**) A cross-sectional fluorescence intensity profile measured along the green dashed bar in **a** with its two peaks fitted to Gaussian functions. (**d**) A cross-sectional intensity profile measured along the green dashed bar in **b** with its two peaks fitted to Gaussian functions. (**e**) An NIR-II image at the level of the gastrocnemius. (**f**) A micro-CT image showing the same area of the limb as in **e**. (**g**) A cross-sectional fluorescence intensity profile measured along the green dashed bar in **e** with its peak fitted to a Gaussian function. (**h**) A cross-sectional intensity profile measured along the green dashed bar in **f** with its peak fitted to a Gaussian function. All scale bars indicate 2 mm.

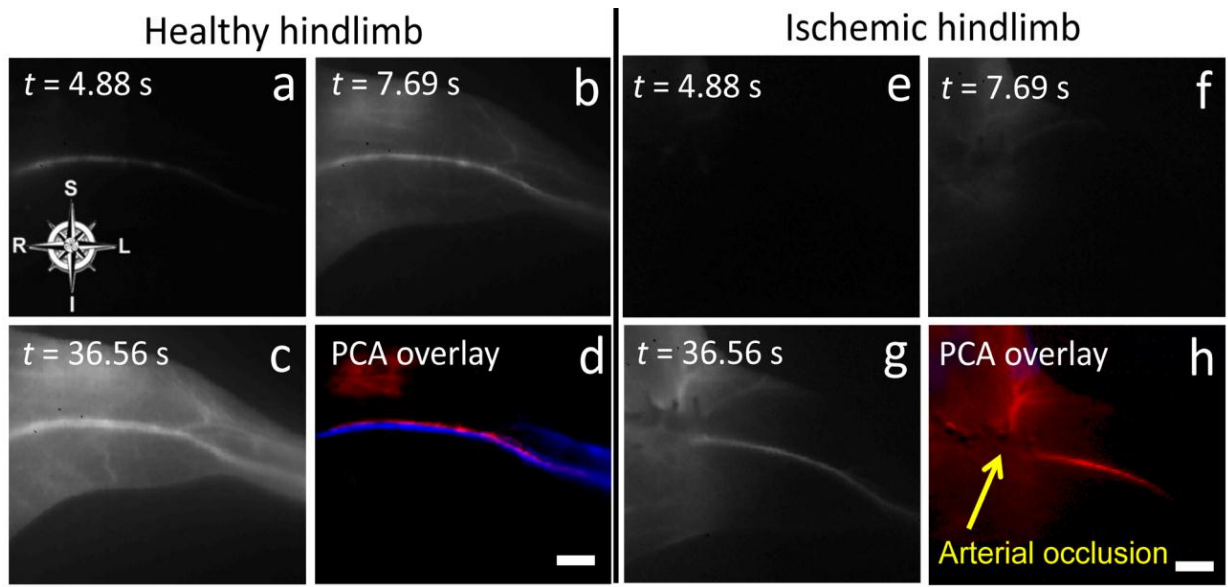

**Figure 3** Differentiation of femoral artery from vein in normal and ischemic mice by NIR-II imaging. (**a–c**) Time course NIR-II fluorescence images of hindlimb blood flow in a control healthy animal. (**d**) PCA overlaid image based on the first 200 frames (37.5 s post injection) of the control animal, where arteries are color-coded in red, while veins are color-coded in blue. (**e–g**) Time course NIR-II fluorescence images of hindlimb blood flow in an ischemic animal. (**h**) PCA overlaid image based on the first 200 frames (37.5 s post injection) of the ischemic animal; only arterial vessels (color-coded in red) can be seen. The level of the experimentally induced arterial occlusion is indicated by the yellow arrow. All scale bars indicate 2 mm.

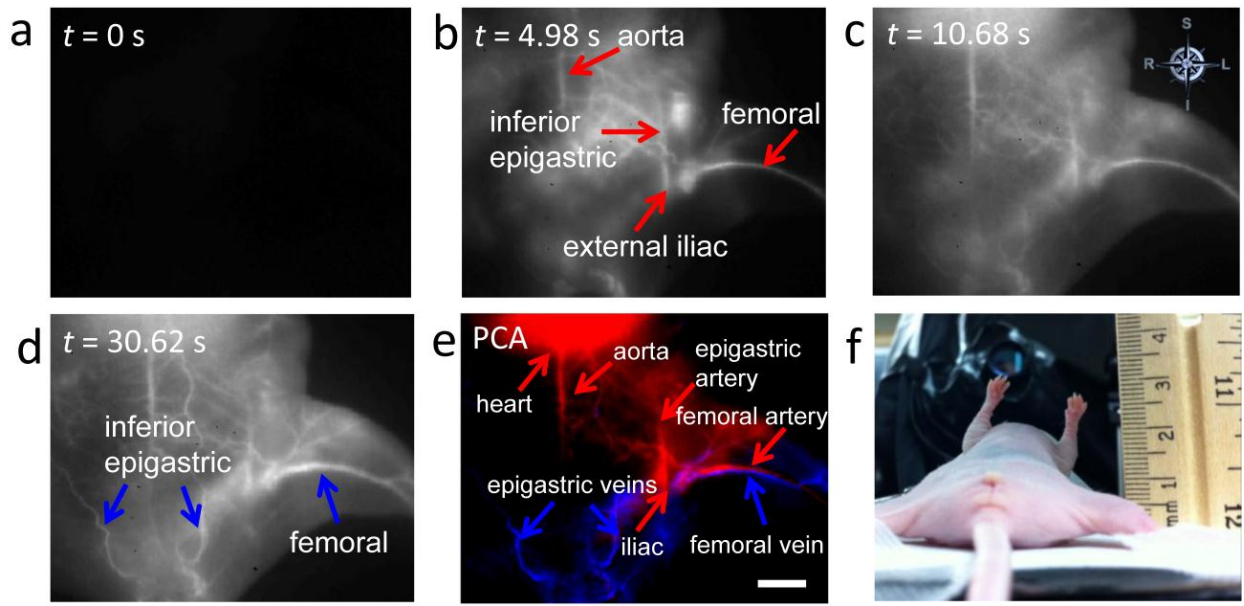

**Figure 4** Differentiation of multiple arterial and venous vessels subserving a larger region of tissue. (**a–d**) Time course NIR-II fluorescence images showing blood flow labeled by SWNT fluorescent tags. (**e**) PCA overlaid image based on the first 170 frames (31.875 s post injection) of the mouse, where arterial vessels are shown in red, while venous vessels are shown in blue. The scale bar indicates 5 mm. (**f**) A digital camera photograph of the mouse imaged in supine position beside a ruler.

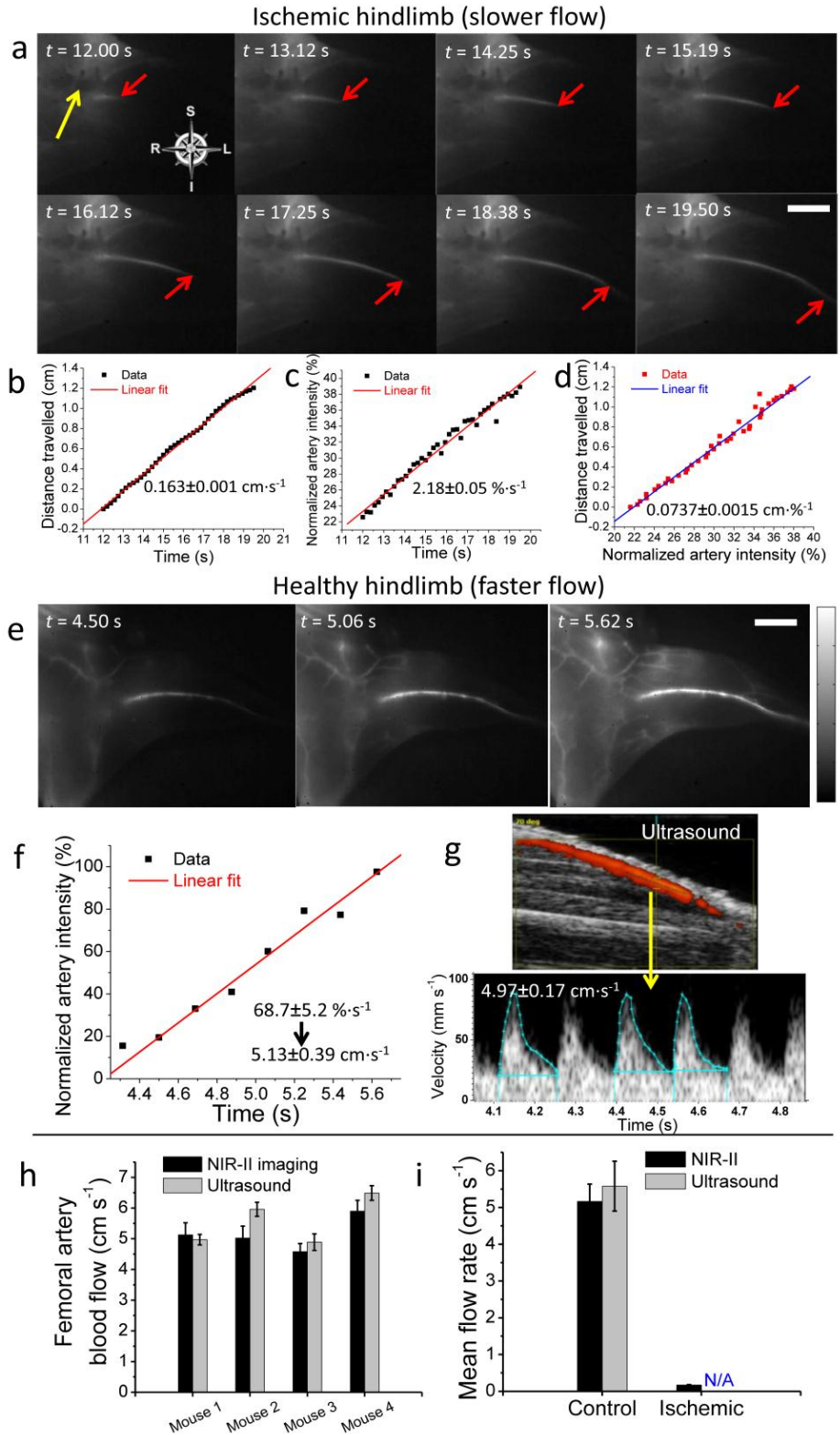

**Figure 5** Femoral artery blood velocity quantification for an ischemic hindlimb and a healthy, control hindlimb by NIR-II imaging and ultrasound. (**a**) Time course NIR-II images showing the flow front

(marked by a red arrow) in the hindlimb of an ischemic mouse, Mouse I1. The yellow arrow indicates the arterial occlusion. (**b**) Distance travelled by the flow front versus time. (**c**) Normalized NIR-II signal in the femoral artery versus time. (**d**) Linear correlation between the artery blood velocity and NIR-II fluorescence increase in femoral artery. (**e**) Time course NIR-II fluorescence images showing the NIR-II intensity increase in the femoral artery of a control mouse, Mouse C1. (**f**) Normalized NIR-II signal in the femoral artery versus time. (**g**) Ultrasound measurement of femoral artery blood velocity in Mouse C1 (top) based on integration of three outlined pulses (bottom). (**h**) Side-by-side comparison of femoral artery velocity obtained from NIR-II method (black) and ultrasonography (gray) for four control healthy mice. Black error bars were based on linear-fitting errors of NIR-II intensity increase plot, while gray error bars reflected the s.d. of three integrated pulses from ultrasonography. (**i**) Average femoral artery blood velocity of control group ($n=4$) and ischemic group ($n=3$), measured by NIR-II method (black) and ultrasound technique (gray). The ischemic blood velocity was not measurable by ultrasonography (shown as 'N/A'). Errors bars reflect the s.d. of each group. Scale bars in **a** and **e** indicate 5 mm, and the intensity scale bar ranges from 0 to 1 for normalized NIR-II fluorescence intensity.

# Tables

**Table 1** Establishing the intensity-to-velocity conversion coefficient for translating NIR-II signal increase to blood velocity. Three ischemic mice, Mouse I1−3, were analyzed based on their blood flow behaviors with SWNTs as NIR-II fluorescence tags.

| Ischemic mouse number | Conversion coefficient [cm·%$^{-1}$] |
|---|---|
| 1 | 0.0737 |
| 2 | 0.0773 |
| 3 | 0.0730 |
| Average | 0.0747±0.0019 |

**Table 2** Dependency of the intensity-to-velocity conversion coefficient on six different variables based on experimental tubing flow results (left column) and numerical simulation (right column), suggesting the coefficient is independent of SWNT concentration, SWNT quantum yield, tubing diameter, ROI length and fluid velocity, while only influenced by the pre-ROI length.

| Variable Control | Experimental Results [cm·%$^{-1}$] | Simulation Results [cm·%$^{-1}$] |
|---|---|---|
| Standard Settings | 0.0778 | 0.0730 |
| Change SWNT concentration | 0.0739 | 0.0730 |
| Change SWNT quantum yield | 0.0756 | 0.0730 |
| Change tubing diameter | 0.0766 | 0.0730 |
| Change ROI length | 0.0789 | 0.0780 |
| Change fluid velocity | 0.0757 | 0.0722 |
| Change pre-ROI length | 0.1648 | 0.1701 |

## Methods

**Preparation of water soluble SWNT-IRDye-800 bioconjugate.** The preparation of water soluble and biocompatible SWNTs can be found in detail in another publication of our group with some modification.[37] In general, raw HiPCO SWNTs (Unidym) were suspended in 1 wt% sodium deoxycholate aqueous solution by 1 hour of bath sonication. This suspension was ultracentrifuged at 300,000 g to remove the bundles and other large aggregates, the supernatant was retained and 0.75 mg·ml$^{-1}$ of DSPE-mPEG(5 kDa) (1,2-distearoyl-*sn*-glycero-3-phosphoethanolamine-N-[methoxy(polyethyleneglycol, 5000)], Laysan Bio) along with 0.25 mg·ml$^{-1}$ of DSPE-PEG(5 kDa)-NH$_2$ (1,2-distearoyl-*sn*-glycero-3-phosphoethanolamine-N-[amino(polyethyleneglycol, 5000)], Sunbright) was added. The resulting suspension was sonicated briefly for 5 min and then dialyzed at pH 7.4 in a 3500 Da membrane (Fisher) with a minimum of six water changes and a minimum of two hours between water changes. To remove aggregates, the suspension was ultracentrifuged again for 1 hour at 300,000 g. This surfactant-exchanged SWNT sample has lengths ranging from 100 nm up to 2.0 μm with the average length of ~500 nm. These amino-functionalized SWNTs were further conjugated with IRDye-800 dye molecule according to the protocol that has been used in our group.[38] Briefly, an SWNT solution with amine functionality at 300 nM after removal of excess surfactant was mixed with 0.1 mM IRDye-800 NHS ester (LI-COR) in PBS at pH 7.4. The reaction was allowed to proceed for 1 h before purification to remove excess IRDye-800 by filtration through 100-kDa filters. The as-made SWNT-IRDye-800 conjugate solution was kept at 4 °C and away from light to avoid photobleaching of IRDye-800 fluorescence.

**UV-Vis-NIR absorption measurements.** UV-Vis-NIR absorption spectrum of the as-made SWNT-IRDye-800 bioconjugate was measured by a Cary 6000i UV-Vis-NIR spectrophotometer, background-corrected for contribution from water. The measured range was 500-820 nm.

**NIR fluorescence spectroscopy of SWNT-IRDye-800 bioconjugate.** NIR fluorescence spectrum was taken using a home-built NIR spectroscopy setup. The excitation source was a 200 W ozone-free mercury/xenon lamp (Oriel), which was dispersed by a monochromator (Oriel) to generate an excitation line with a central wavelength of 785 nm and a bandwidth of 15 nm. The excitation light was allowed to pass through the solution sample in a 1 mm path cuvette (Starna Cells, Inc.) and the emission was collected in a transmission geometry. The excitation light was rejected using a 790-nm long-pass filter (Semrock) so that the fluorescence of both IRDye-800 and SWNTs could be collected in the 790-1500 nm emission range. The emitted light was directed into a spectrometer (Acton SP2300i) equipped with a liquid-nitrogen-cooled InGaAs linear array detector (Princeton OMA-V). Spectra were corrected post-collection to account for the sensitivity of the detector and extinction feature of the filter using the MATLAB software.

**Determination of cytotoxicity of SWNTs.** We determined the SWNT toxicity *in vitro* by MTS assay using a CellTiter96 kit (Promega) on human dermal microvascular endothelial cells (Lonza). ~5000 cells were incubated per well with 100 μl of EGM2MV growth media (Lonza) and serially diluted SWNT solution (*n*=3 for each concentration). The cells were kept at 37 °C in a humidified atmosphere containing 5% $CO_2$ for 24 hours in the presence of SWNTs at different concentrations. Immediately before addition of 15 μL of CellTiter96, a colorimetric indicator of cell viability, the SWNT spiked medium was removed from each well plate and replaced with fresh medium. This prevented any interference in the absorbance readings from SWNTs. After 1 hour, the color change was quantified using a plate reader and taking absorbance readings at 490 nm. Cell viability was plotted as a fraction of the absorbance of control wells incubated without SWNTs.

**Mouse handling, surgery and injection.** 6-week-old female athymic nude mice were obtained from Charles River. All animal studies were approved by Stanford University's Administrative Panel on Laboratory Animal Care. Induction of unilateral hindlimb ischemia was performed according to our previous studies.[28,30,39] Control, unsurgerized mice (*n*=4) and mice with induced ischemia (*n*=3) were used in the study. For the injection of nanotube solution, a 28 gauge syringe needle was inserted into the lateral tail vein, allowing for bolus injection during the first frames of imaging. All mice were initially anesthesized before imaging in a knockdown box with 2 L·min$^{-1}$ $O_2$ gas flow mixed with 3% Isoflurane. A nose cone delivered 1.5 L·min$^{-1}$ $O_2$ gas and 3% Isoflurane throughout imaging.

*In vivo* **NIR fluorescence imaging with tunable magnifications.** Animals were mounted on a heated stage in the supine position beneath the laser at 10 min post injection. NIR fluorescence images were collected using a 1344 × 1024 pixel silicon CCD camera (Hamamatsu) for collecting photons in NIR-I and a liquid-nitrogen-cooled, 320 × 256 pixel two-dimensional InGaAs array (Princeton Instruments) for collecting photons in NIR-II. A flip mirror was used to switch photon collection between the two cameras (**Fig. 1d**). The excitation light was provided by a 785-nm diode laser (Renishaw) coupled to a 4.5-mm focal length collimator (Thorlabs) and filtered by a 790-nm bandpass filter with a bandwidth of 10 nm (Thorlabs). The excitation power density at the imaging plane was 8 mW·cm$^{-2}$, much lower than the safe exposure limit of 296 mW·cm$^{-2}$ at 785 nm determined by the International Commission on Non-ionizing Radiation Protection.[40] The emitted light from the animal was filtered through a 790-nm long-pass filter (Semrock) and an 850-nm short-pass filter (Thorlabs) coupled with the Si camera for the NIR-I imaging window, or through a 900-nm long-pass filter (Thorlabs) and an 1100-nm long-pass filter (Thorlabs) coupled with the InGaAs camera for NIR-II imaging. A lens set was used for obtaining tunable magnifications, ranging from 1× to 2.5× magnification by changing the relative position of two NIR achromats (200 mm and 75 mm, Thorlabs), and from 2.5× to 7× magnification by changing the relative position of two other NIR achromats (150 mm and 200 mm, Thorlabs). A binning of 4 and an exposure time of 300 ms were used for the Si camera (1344 × 1024 pixels) to capture images in the NIR-I window, and a binning of 1 and an exposure time of 300 ms were used for the InGaAs camera (320 × 256 pixel) to capture images in the NIR-II window. Different binnings were used to compensate the difference of array size of the two cameras.

**Microscopic computed tomography for vascular imaging.** Micro-CT scans were performed using a MicroCAT II micro-CT scanner (Siemens Preclinical Solutions) using the following parameters: X-ray voltage 80 kVp, anode current 50 mA, and exposure time 2000 ms per 576 frames through 360° rotation. Mice were injected with blood pool contrast agent Fenestra VC (50 mg·ml$^{-1}$ iodine, Advanced Research Technologies) intravenously into the lateral tail vein at 0.3 ml·(20 g)$^{-1}$ body weight with a 28-gauge needle. Animals ($n$=2) were scanned 1 h post-injection for 1 h total scan time. Three-dimensional reconstruction was performed by COBRA 1.5 and visualized using Amira 5.4. Resulting voxel dimension was 40 μm.

**Video-rate imaging in the NIR-II window.** Video-rate imaging was performed on the same homebuilt imaging system as in the steady-state imaging case except that only an InGaAs camera was used for imaging in the NIR-II window. The excitation light was provided by an 808-nm diode laser (RMPC lasers) coupled to a 4.5-mm focal length collimator (Thorlabs) and filtered by an 850-nm short-pass filter and a 1000-nm short-pass filter (Thorlabs). The excitation power density at the imaging plane was 140 mW·cm$^{-2}$, lower than the safe exposure limit 329 mW·cm$^{-2}$ at 808 nm.[40] The emitted light from the animal was filtered through a 900-nm long-pass filter and an 1100-nm long-pass filter (Thorlabs) so that the intensity of each pixel in the InGaAs 2D array represented light in the 1.1~1.7 μm range. A lens pair consisting of two achromats (200 mm and 75 mm, Thorlabs) was used focus the image onto the detector with a magnification of 2.5×. The InGaAs camera was set to expose continuously, and NIR-II fluorescence images were acquired with LabVIEW software. The exposure time for all images shown in the videos was 100 ms. There was an 87.5-ms overhead in the readout, leading to an average time of 187.5 ms between consecutive frames and a frame rate of 5.3 frames·s$^{-1}$.

**Dynamic contrast-enhanced imaging based on PCA.** Dynamic contrast-enhanced images were obtained in a similar fashion to previous work by the Hillman group[23] and our group.[8] First 200 consecutive frames immediately after injection were loaded into an array using MATLAB software, and the built-in *princomp* function was used to perform PCA. Empirically features showing up early in the video are grouped in the negative fourth principal component and features showing up later in the video are grouped in the negative second principal component. Therefore negative pixels for the second principal component were color-coded in blue to represent venous vessels while negative pixels for the fourth principal component were color-coded in red to represent arterial vessels.

**Quantification of blood velocity and flow based on NIR-II fluorescence.** Average region of interest (ROI) NIR-II fluorescence intensity was computed using MATLAB software within a given arterial region of each frame, which was determined by PCA analysis as explained in previous Methods sections. Then the NIR-II fluorescence intensity was plotted as a function of time to reveal the change over 4 min after intravenous injection. The intensity in the femoral artery increased rapidly within the first 5-10 s before residing. The NIR-II intensity values were normalized against the maximum intensity within the ROI time traces to compensate for any differences in actual injection dose and relative fluorescence quantum yield. Linear fit was then performed on the rising edge of the normalized plot, and its slope (in $\%\cdot s^{-1}$) was translated into blood velocity in terms of $cm\cdot s^{-1}$ by using the intensity-to-velocity conversion coefficient. Since the NIR-II signal increase was averaged over the ROI, the velocity value was the mean blood flow velocity along the femoral artery. Blood flow (F) in $ml\cdot min^{-1}$ ($cm^3\cdot min^{-1}$) was calculated based on the blood velocity (V) in $cm\cdot s^{-1}$ and the diameter of the vessel (d) in μm as follows: $F = \pi*(d/2)^2 * V * 6*10^{-7}$.

**Validation of intensity-to-velocity coefficient based on tubing flow model.** A solution of SWNTs with known concentration was pumped by a syringe pump into catheter tubing with known diameter filled with purified water at a preset velocity (**Fig. S5k**). The following settings were used for the standard condition: SWNT concentration = 0.10 $mg\cdot ml^{-1}$, SWNT fluorescence quantum yield (QY) ~ 2.5%, tubing diameter = 760 μm, tubing length within ROI (i.e., ROI length) = 2.5 cm, fluid velocity = 1.4 $cm\cdot s^{-1}$ and tubing length before ROI (i.e., pre-ROI length) = 8.5 cm. To screen the dependency of the coefficient on all these variables, the parameters were changed one at a time while keeping all others unchanged: SWNT concentration changed to 0.025 $mg\cdot ml^{-1}$, SWNT fluorescence QY changed to 5.0%, tubing diameter changed to 380 μm, ROI length changed to 1.25 cm, fluid velocity changed to 0.14 $cm\cdot s^{-1}$ and pre-ROI length changed to 19.5 cm. Under each combination of settings, the NIR-II fluorescence intensity within the selected ROI was plotted as a function of time from immediately after injection to the time when intensity plateaued. Then NIR-II intensity values were normalized against the maximum intensity (plateau intensity) within the ROI time traces. Linear fit was then performed on the linear rise of each normalized plot to obtain a slope in $\%\cdot s^{-1}$. The intensity increase rate (slope) in $\%\cdot s^{-1}$ was then used to divide the velocity in $cm\cdot s^{-1}$ to obtain the intensity-to-velocity conversion coefficient in $cm\cdot\%^{-1}$. For numerical simulation, a tubing linear flow model with axial mixing was adapted from a previous publication.[29] A sigmoidal function with time and velocity dependence was used to simulate NIR-II intensity distribution at the flow front upon mixing using MATLAB software. The flow front function $F(x,v,t)$ was given analytically by:

$$F(x,v,t) = \frac{I \cdot \varepsilon cd \cdot QY}{1 + e^{\frac{x-vt}{A_0 + Kvt}}}$$

where the excitation power density $I$, absorption coefficient $\varepsilon$, initial degree of mixing $A_0$ (~0.001cm) and mixing constant $K$ (~0.5) are fixed, while concentration $c$, tubing diameter $d$, fluorescence quantum yield $QY$ and velocity $v$ were varied in the simulations to find the dependency of the coefficient. And the normalized ROI intensity was computed numerically as follows,

$$I_{norm}(t) = \frac{\int_{x=L_{\text{pre-ROI}}}^{x=L_{\text{pre-ROI}}+L_{\text{ROI}}} F(x,v,t)dx}{\int_{x=L_{\text{pre-ROI}}}^{x=L_{\text{pre-ROI}}+L_{\text{ROI}}} F(x,v,+\infty)dx}$$

which was then plotted against time (**Fig. S5l-r**). The linear rise region was fit to a linear equation, and the slope was used to divide the velocity, giving the intensity-to-velocity conversion coefficient:

$$Coeff. = \frac{v}{\partial I_{norm}(t) / \partial t}$$

**Ultrasound for quantifying blood flow in femoral artery.** Ultrasound measurements were performed using a linear real-time transducer (40 MHz) connected to a Vevo 2100 ultrasound system (VisualSonics). The femoral artery was identified employing Duplex-ultrasonography (B-Mode and power Doppler). Flow velocity profiles were recorded by cw-Doppler imaging. Velocity-time integrals (VTI) and cardiac cycle length (CL) were measured using the Vevo 2100 device software. Arterial diameter (d) was known for each animal from previous NIR-II imaging. Femoral flow (F) was calculated as: F = Stroke volume (SV)* heart rate (HF) = arterial cross-sectional area (CSA) * VTI * HF = $\pi*(d/2)^2$ * VTI * 60,000 / CL and averaged from three cardiac cycles.

## References for Methods

# Supplementary Information for

**Multifunctional *in vivo* vascular imaging using near-infrared II fluorescence**

Guosong Hong[1,3], Jerry C. Lee[2,3], Joshua T. Robinson[1], Uwe Raaz[2], Liming Xie[1], Ngan F. Huang[2], John P. Cooke[2] & Hongjie Dai[1]

[1] Department of Chemistry, Stanford University, Stanford, California 94305, USA

[2] Division of Cardiovascular Medicine, School of Medicine, Stanford University, Stanford, California 94305, USA

[3] These authors contribute equally to this work.

# Supplementary Figures

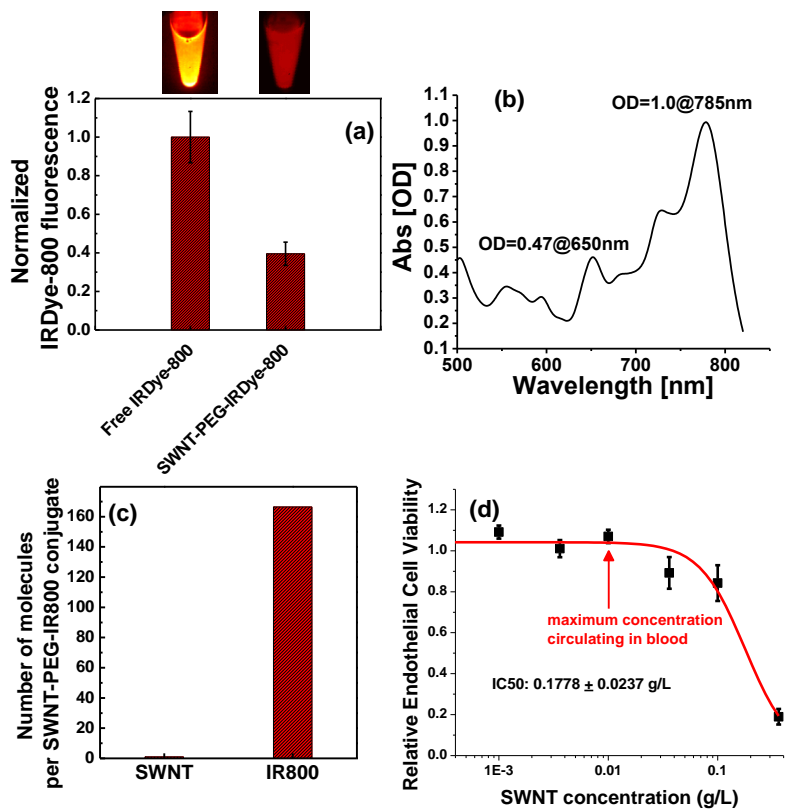

**Figure S1.** Chemical composition and cytotoxicity of SWNT-PEG-IRDye-800 conjugate. (**a**) Normalized IRDye-800 fluorescence for free IRDye-800 and SWNT-PEG-IRDye-800 at the same concentration of IRDye-800, suggesting the photoluminescence of IRDye-800 was quenched by ~60% due to the attachment to SWNTs through the PEG-chains. Inset pictures on top are corresponding photoluminescence images taken in the NIR-I window. (**b**) UV-Vis-NIR absorption spectrum of SWNT-PEG-IRDye-800 conjugate taken in a 1 mm cuvette, where the OD values at 650 nm and 785 nm are used for calculating the average number of SWNT and IRDye-800 molecules in SWNT-PEG-IRDye-800 conjugate. (**c**) Calculated number of SWNT and IRDye-800 molecules in each SWNT-PEG-IRDye-800 conjugate, indicating on average 167 IRDye-800 molecules on each SWNT backbone. The large number of IRDye-800 molecules attached to SWNTs ensured sufficient NIR-I emitters despite fluorescence quenching of IRDye-800 of ~60% in **a**. (**d**) Determination of half maximal inhibitory concentration (IC50) of SWNTs for endothelial cells. Original data (black squares) were fitted to sigmoidal function, revealing an IC50 value of 0.1778±0.0237 g/L. Errors bars reflect the standard deviation of relative endothelial cell viability values from 3 wells of cells incubated at the same SWNT concentration.

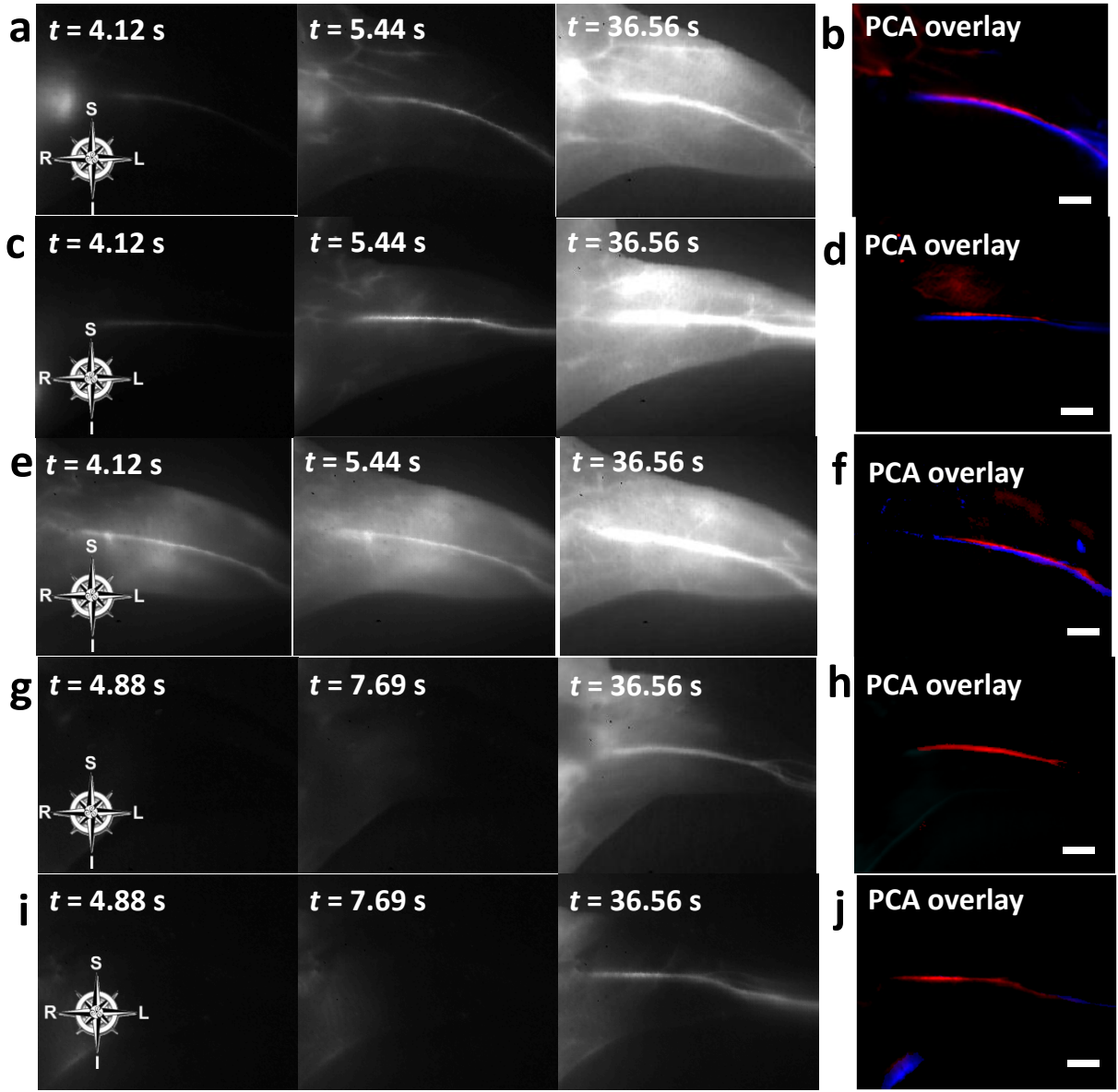

**Figure S2.** Differentiation of femoral artery and vein inside healthy, control hindlimbs of Mouse C2–4 (**a–f**) and ischemic hindlimbs of Mouse I2–3 (**g–j**). (**a,c,e,g,i**) Time course NIR-II fluorescence images showing hindlimb blood flow labeled by SWNT fluorescent tags. (**b,d,f,h,j**) PCA overlaid images based on the first 200 frames (37.5 s post injection) of each mouse, where arterial vessels are shown in red, while venous vessels are shown in blue. The scale bars all indicate 2 mm.

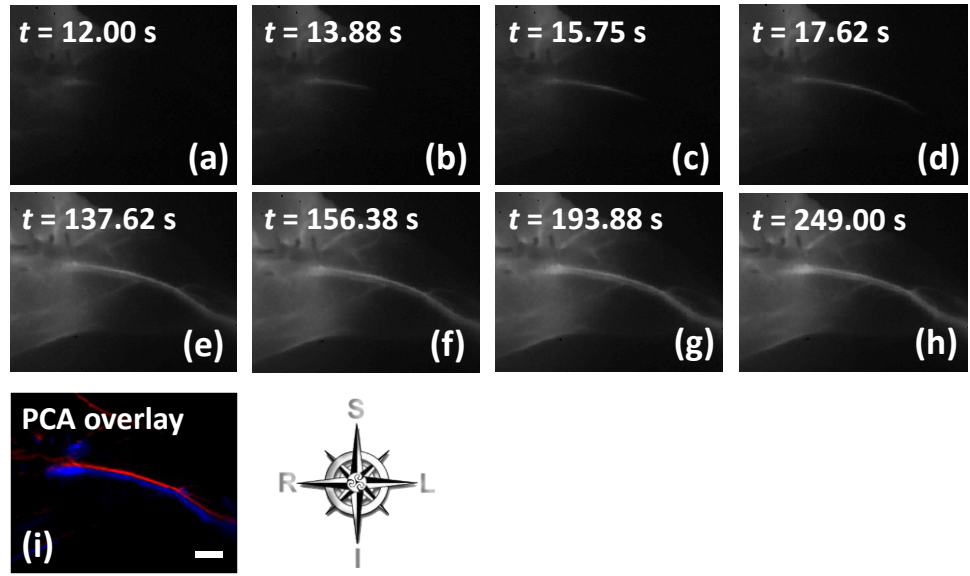

**Figure S3.** Differentiation of femoral artery and vein inside an ischemic hindlimb of Mouse I1 based on a longer post-injection imaging time. (**a–h**) Time course NIR-II fluorescence images showing hindlimb blood flow labeled by SWNT fluorescent tags. Note that femoral vein does not show up until after 2 min due to reduced blood flow of ischemia. (**i**) PCA overlaid image based on 1320 frames (up to 247.5 s post injection) of the mouse, where arterial vessels are shown in red, while venous vessels are shown in blue. The scale bar indicates 2 mm.

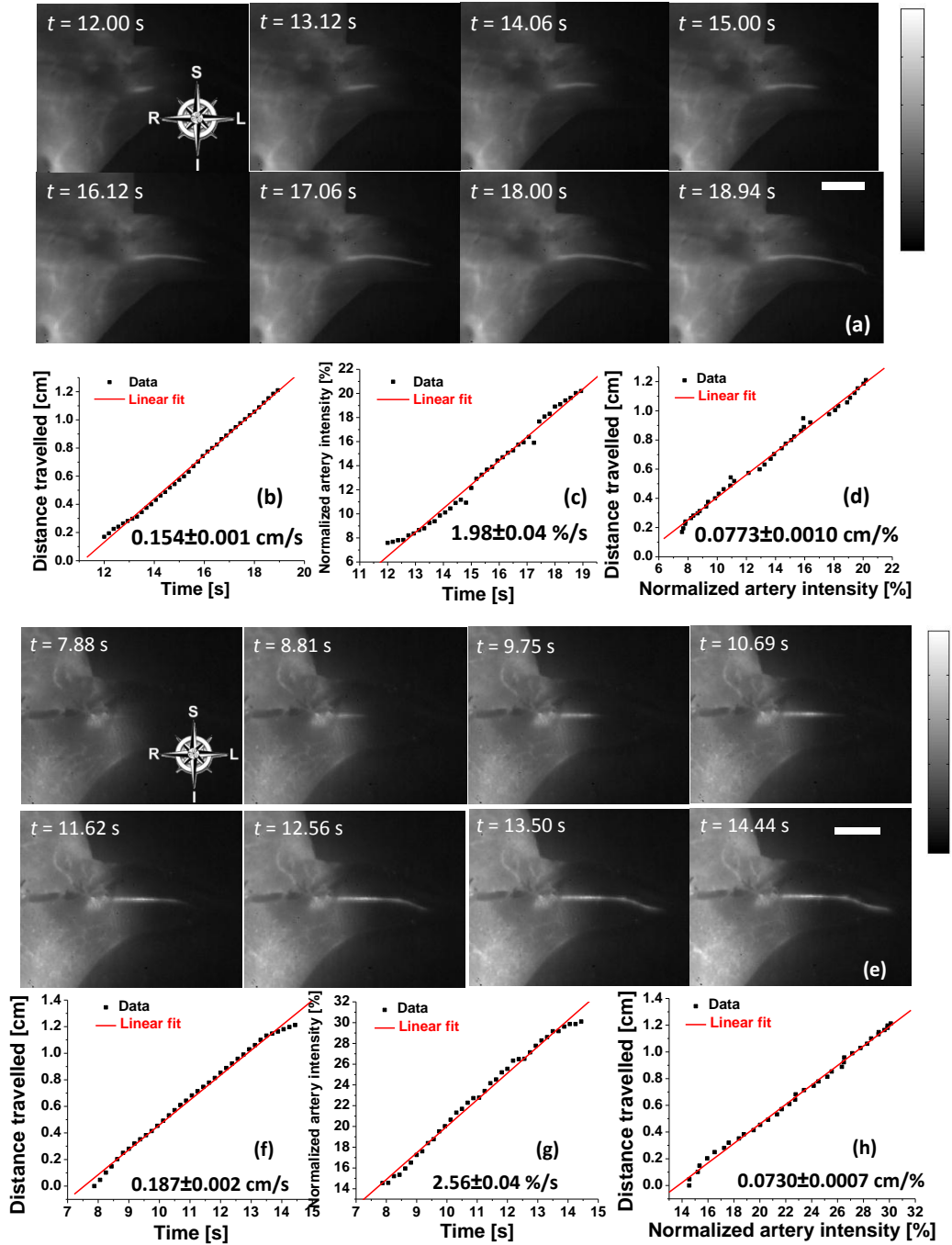

**Figure S4.** Arterial blood velocity analysis on the hindlimbs of ischemic Mouse I2 (**a-d**) and I3 (**e–h**). (**a,e**) NIR-II fluorescence images showing the flow front labeled by SWNT fluorescent tags. (**b,f**) Distance travelled by the flow front plotted as a function of time. (**c,g**) Normalized NIR-II signal in the femoral artery plotted as a function of time. (**d,h**) Linear correlation between the artery blood velocity and NIR-II fluorescence increase in the corresponding artery area. Scale bars in **a,e** indicate 5 mm and the intensity scale bars range from 0 to 1.

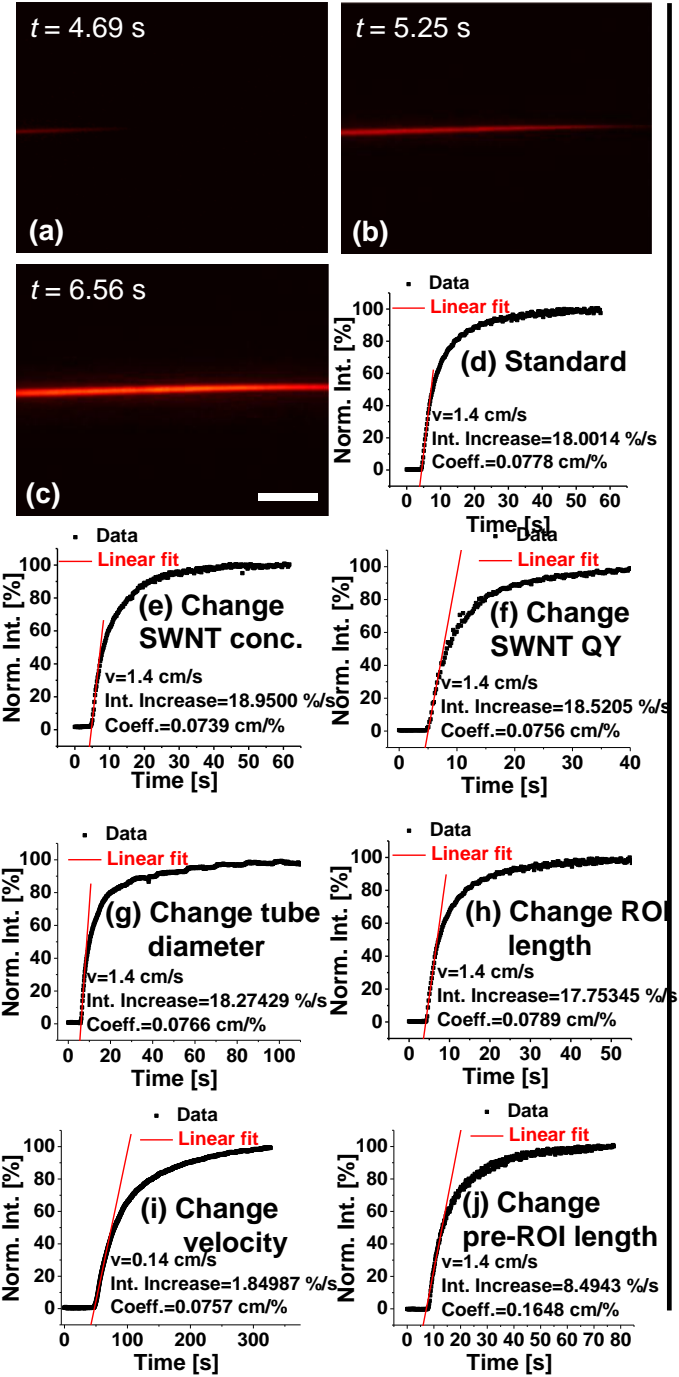
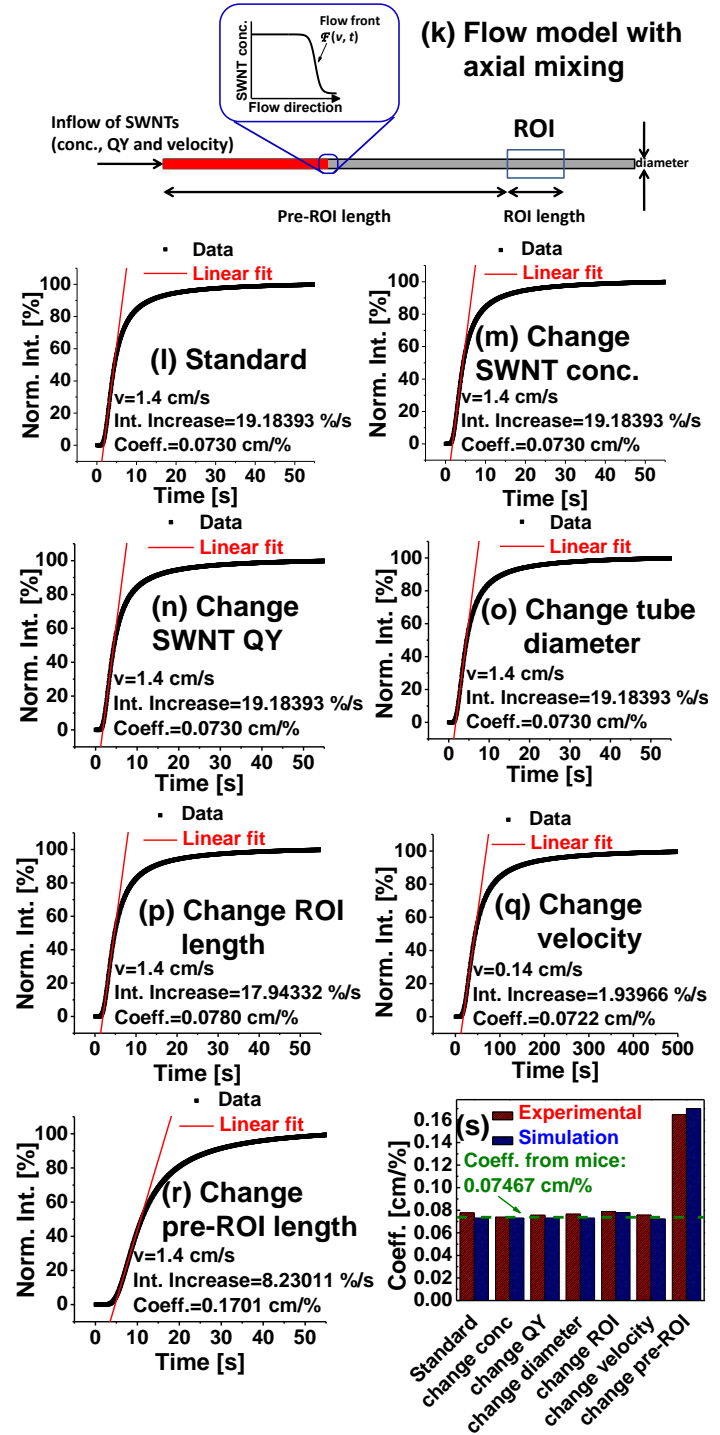

**Figure S5.** Variable dependency study of the intensity-to-velocity conversion coefficient based on both experimental (**a-j**) and simulation (**l-r**) results of the tubing flow setup (**k**). To find which variable(s) influence(s) the value of the coefficient, six possible parameters are changed one at a time from the standard settings while keeping all other parameters unchanged. The standard settings are given by: SWNT concentration = 0.10 mg/mL, SWNT fluorescence quantum yield (QY) = 2.5%, tubing diameter = 760 μm, tubing length within ROI (i.e., ROI length) = 2.5 cm, fluid velocity = 1.4 cm/s and tubing length before ROI (i.e., pre-ROI length) = 8.5 cm. (**a-c**) NIR-II fluorescence images of SWNTs flowing through the tubing at 4.69 s, 5.25 s and 6.56 s post injection. Scale bar indicates 5 mm. (**d-j**) Experimental normalized NIR-II intensity increase curve as a function of time when: at the standard settings (**d**), only the SWNT concentration is reduced by 4 times (**e**), only the QY of SWNT is increased by 2 times (**f**), only the tubing diameter is reduced by 2 times (**g**), only the ROI length is reduced by 2 times (**h**), only the fluid velocity is reduced by 10 times (**i**), and only the pre-ROI is increased by 2 times (**j**). A linear fit based on the onset increase in each case gives a coefficient of 0.0778 cm/%, 0.0739 cm/%, 0.0756 cm/%, 0.0766 cm/%, 0.0789 cm/%, 0.0757 cm/% and 0.1648 cm/%, respectively. Note the coefficient remains invariant except in the case of a change in the pre-ROI length. (**k**) A schematic drawing of the tubing flow experiment setup, where all six parameters are labeled. The same setup is used in the numerical simulation, and the flow front is simulated by a sigmoidal function, the shape of which is dependent of both time and fluid velocity, based on a tubing flow model with axial mixing.[1,2] (**l-r**) Simulational normalized NIR-II intensity increase curve as a function of time when at the standard settings (**l**), only the SWNT concentration is reduced by 4 times (**m**), only the QY of SWNT is increased by 2 times (**n**), only the tubing diameter is reduced by 2 times (**o**), only the ROI length is reduced by 2 times (**p**), only the fluid velocity is reduced by 10 times (**q**), and only the pre-ROI is increased by 2 times (**r**). A linear fit based on the onset increase in each case gives a coefficient of 0.0730 cm/%, 0.0730 cm/%, 0.0730 cm/%, 0.0730 cm/%, 0.0780 cm/%, 0.0722 cm/% and 0.1701 cm/%, respectively. Note the coefficient remains invariant except in the case of a change in the pre-ROI length. (**s**) A bar chart summarizing the conversion coefficients derived from experimental results (red bars) and simulation (blue bars), with comparison to the coefficient derived from the femoral arterial flow in mice (green dashed line), showing the conversion coefficient invariant of SWNT concentration, SWNT quantum yield, tubing diameter, ROI length and fluid velocity, while only influenced by the pre-ROI length.

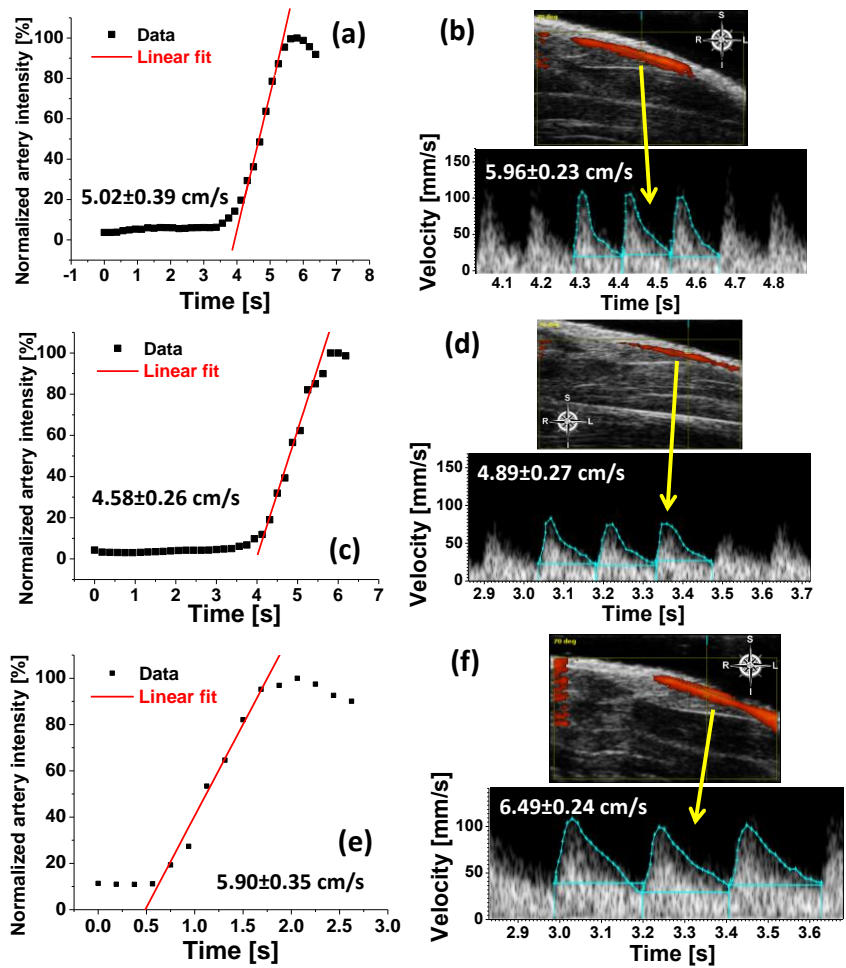

**Figure S6.** Reproduced femoral artery blood flow quantification in three healthy, control nude mice (Mouse C2–4). (**a**) Normalized NIR-II signal in the femoral artery plotted as a function of time, revealing a blood velocity of 5.02±0.39 cm/s after conversion for Mouse C2. (**b**) Power Doppler of femoral artery blood flow in Mouse C2 shown in the top graph, revealing an average blood velocity of 5.96±0.23 cm/s, based on the velocity time integral (VTI) of three cardiac cycles as shown by pulsed wave Doppler in the bottom graph. (**c**) Normalized NIR-II signal in the femoral artery plotted as a function of time, revealing a blood velocity of 4.58±0.26 cm/s after conversion for Mouse C3. (**d**) Power Doppler of femoral artery blood flow in Mouse C3 shown in the top graph, revealing an average blood velocity of 4.89±0.27 cm/s, based on the VTI of three cardiac cycles as shown by pulsed wave Doppler in the bottom graph. (**e**) Normalized NIR-II signal in the femoral artery plotted as a function of time, revealing a blood velocity of 5.90±0.35 cm/s after conversion for Mouse C4. (**f**) Power Doppler of femoral artery blood flow in Mouse C4 shown in the top graph, revealing an average blood velocity of 6.49±0.24 cm/s, based on the VTI of three cardiac cycles as shown by pulsed wave Doppler in the bottom graph.

# References for SI